\documentclass[aps,prb,reprint,superscriptaddress]{revtex4-1}
\usepackage{epsfig,amssymb, amsmath, amsthm,graphicx,color,hyperref}
\def\bfr{{\bf r}}

\begin{document}

\title{Quantum many-body models with cold atoms coupled to photonic crystals}

\author{J.~S.~Douglas}
\email[email address: ]{james.douglas@icfo.eu}
\affiliation{ICFO-Institut de Ciencies Fotoniques, 08860 Castelldefels, Barcelona, Spain}
\author{H.~Habibian}
\affiliation{ICFO-Institut de Ciencies Fotoniques, 08860 Castelldefels, Barcelona, Spain}
\author{C.-L. Hung}
\altaffiliation{Present address: Department of Physics and Astronomy, Purdue University, West Lafayette, Indiana 47907, USA}
\affiliation{Norman Bridge Laboratory of Physics and Institute for Quantum Information and Matter, California Institute of Technology, Pasadena, California 91125, USA}
\author{A.~V.~Gorshkov}
\affiliation{Joint Quantum Institute and Joint Center for Quantum Information and Computer Science, NIST/University of Maryland, College Park, Maryland 20742, USA}
\author{H.~J.~Kimble}
\affiliation{Norman Bridge Laboratory of Physics and Institute for Quantum Information and Matter, California Institute of Technology, Pasadena, California 91125, USA}
\author{D.~E.~Chang}
\affiliation{ICFO-Institut de Ciencies Fotoniques, 08860 Castelldefels, Barcelona, Spain}

\date{\today}

\begin{abstract}
Using cold atoms to simulate strongly interacting quantum systems represents an exciting frontier of physics. However, as atoms are nominally neutral point particles, this limits the types of interactions that can be produced. We propose to use the powerful new platform of cold atoms trapped near nanophotonic systems to extend these limits, enabling a novel quantum material in which atomic spin degrees of freedom, motion, and photons strongly couple over long distances. In this system, an atom trapped near a photonic crystal seeds a localized, tunable cavity mode around the atomic position. We find that this effective cavity facilitates interactions with other atoms within the cavity length, in a way that can be made robust against realistic imperfections. Finally, we show that such phenomena should be accessible using one-dimensional photonic crystal waveguides in which coupling to atoms has already been experimentally demonstrated.
\end{abstract}

\maketitle

Trapped ultracold atoms are a rich resource for physicists. Isolated from the environment and routinely manipulated, they can act as a quantum simulator for a wide variety of physical models \cite{Bloch2012a}. However, while short-range interactions between atoms can be adjusted by Feshbach resonance, these systems typically lack the long-range interactions required to produce some of the most interesting condensed matter phenomena. For example, exotic phases such as supersolids are predicted in systems with long-range interactions \cite{Batrouni1995a}, as well as Wigner crystallization \cite{Wigner1934} and topological states \cite{Micheli2006a}. Long-range interactions can also lead to the breakdown of concepts such as additivity in statistical mechanics \cite{Campa2009a,Shahmoon2014a} and the violation of speed limits (Lieb-Robinson bounds) for the propagation of information \cite{Hauke2013a,Richerme2014a,Jurcevic2014a}. 
As a result, there are active efforts to achieve long-range interactions using specific properties of the atoms \cite{Lahaye2009a}, such as large magnetic moment \cite{Griesmaier2005a,Lu2011a}, Rydberg excitation \cite{Saffman2010a}, or using polar molecules \cite{Ni2008a}. 

We investigate another paradigm, where instead of relying on the atomic properties, we design the medium via which the atoms interact. Specifically, by coupling the atoms via the photon modes of a photonic crystal. Our proposal is inspired by demonstrations of strong coupling of photons in nanophotonic systems with individual solid-state emitters \cite{Khitrova2006a} and more recently with cold atoms \cite{Vetsch2010a,Goban2012a,Thompson2013a,Goban2013a}. For example, systems of $\sim 10^3$ atoms have been trapped by and coupled to the evanescent guided modes of nanofibers~\cite{Vetsch2010a,Goban2012a}, and single atoms have been coupled to photonic crystal cavities~\cite{Thompson2013a} and waveguides~\cite{Goban2013a}. One aim of these efforts is 
An early motivation for these efforts was the aim to utilize strong
An original motivation of these efforts was to utilize strong, controlled light-matter interactions for quantum information processing and networks~\cite{Kimble2008a}. Here, we show that atoms interfaced with photonic crystals can also have remarkable consequences for the exploration of quantum many-body physics~\cite{Greentree2006a,Hartmann2006a,Angelakis2007a}.

A photonic crystal is a periodic dielectric structure that controls the propagation of light~\cite{Joannopoulos2008}. By introducing a defect into this regular structure cavity modes for the light may be induced. In this work, we demonstrate that a single atom trapped near an otherwise perfect photonic crystal can also seed a localised cavity mode around the atom. The physics of the atom coupled with the photonic crystal can then be understood by a direct mapping to cavity quantum-electrodynamics (QED) allowing intuition and results to be carried from this well-developed field.
When many atoms are trapped, these dynamically induced cavities mediate coherent interactions between atoms \cite{Kurizki1990a,John1991a}. 
These interactions can extend over distances of the order of 100 optical wavelengths and we describe here for the first time the principles of how this long-range coupling between atoms can be achieved and tuned in the framework of current experiments.

In particular, relative to prior work in this area \cite{John1990a,Kurizki1990a,John1991a,Bay1997a,Lambropoulos2000a,Shahmoon2013a,Shahmoon2014a}, we show how the type of effective spin interaction (with spin encoded in atomic internal degrees of freedom) and its spatial range can be engineered and dynamically tuned using available atomic structures and external controlling laser fields. Furthermore, we validate in detail the limits of applicability of our theoretical model, comparing with full numerical simulation (Greens function calculations) of an actual one-dimensional photonic crystal waveguide used in experiments. Importantly we provide realistic descriptions of the fields and atoms, taking into account photon loss and localization of photons due to imperfections in the dielectric structure and loss resulting from free-space atomic emission.

The principles that we elaborate have broad applicability to atom-photon interactions in nano-photonics, including significantly to two dimensional photonic crystal geometries \cite{Tudela2014a}. For definiteness, here we focus on the implementation in one-dimensional waveguides, and show how such phenomena should be accessible using photonic crystal waveguide geometries in which coupling to atoms has already been experimentally demonstrated \cite{Goban2013a,Yu2013a}. More generally, our work provides a platform to realize new regimes of physics involving simultaneous strong and long-range coupling between spins, phonons, and photons, enabled by the strong atom-light interactions possible in the nanophotonic system.


\begin{figure}
\centering
\includegraphics{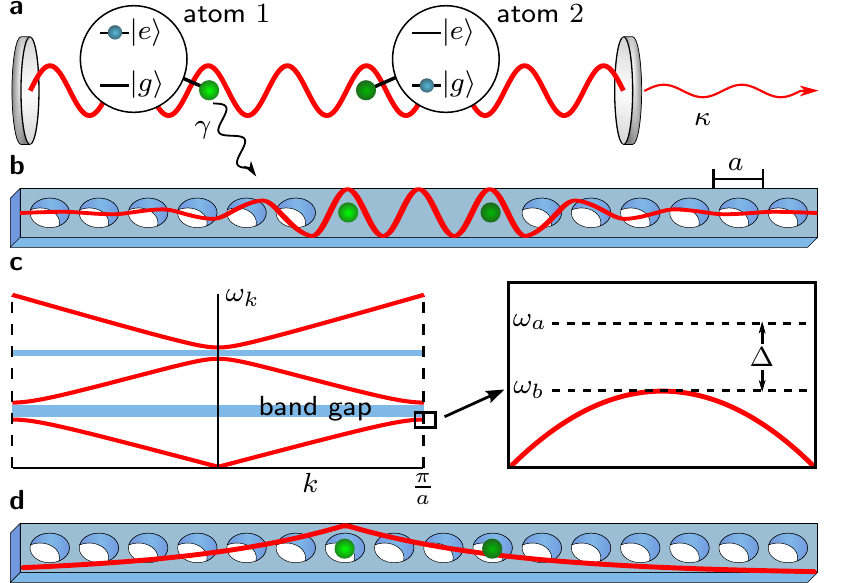}
\caption{\textbf{From cavity-QED to atom-induced cavities in photonic crystals} (a) Two atoms are coupled with strength $g_c$ to a single mode of a Fabry-Perot cavity, enabling an excited atom (atom 1) to transfer its excitation to atom 2 and back. The coherence of this process is reduced by the cavity decay (rate $\kappa$) and atomic spontaneous emission into free space (rate $\gamma$).
(b) Photonic crystals are alternating dielectric materials, shown here as oval air holes in a dielectric waveguide, with unit cell length $a$. A defect, such as caused by removing or altering the hole sizes, can lead to a localized photonic mode (red).
Atoms coupled to such a system may then interact via this mode in an analogous manner to in (a). (c) A typical band structure of a 1d photonic crystal, illustrating the guided mode frequency $\omega_k$ versus the Bloch wavevector $k$ in the first Brillouin zone. We are interested in the case where atoms coupled to the crystal have resonance frequency $\omega_a$ close to the band edge frequency $\omega_b$, with $\Delta\equiv \omega_a-\omega_b$.
(d) An atom near a photonic crystal can act as a defect, inducing its own cavity mode with an exponentially decaying envelope (red). A second atom can couple to this mode to produce an interaction similar to in (a) and (b), but where the strength now depends on the inter-atomic distance.}
\label{fig:cavity_prog}
\end{figure}


\section*{Effective long range interactions}

Long-range interactions between particles often occur through the exchange of photons.
A simple example consists of two-level atoms with transition frequency $\omega_a = 2\pi c/\lambda$ interacting via a single mode of a Fabry-Perot cavity that has resonance frequency $\omega_c$, as shown in Fig.~\ref{fig:cavity_prog}a. Momentarily neglecting losses, a single atom in the cavity
is described by
the Jaynes-Cummings model \cite{Agarwal1995},
$H = \hbar \omega_a \sigma_{ee}+ \hbar\omega_c \hat{a}^\dagger \hat{a}+\hbar g_c (\sigma_{eg} \hat{a} +\mbox{h.c.})$,
where $\sigma_{\mu\nu} = |\mu\rangle\langle\nu|$ operate on the internal atomic state and the cavity mode excitation has associated annihilation operator $\hat{a}$. The coupling between the atom and the cavity mode $g_c = d_{eg}\sqrt{\omega_c/(2\hbar\epsilon_0 V)}$ depends on the strength of the dipole matrix element $d_{eg}$ of the two level transition and on the cavity mode volume $V$.
For a single excitation, the eigenstates are dressed states --- superpositions of the excitation being purely atomic and purely in the cavity mode --- given by $|\psi_1\rangle = \cos\theta |e\rangle|0\rangle + \sin\theta |g\rangle|1\rangle$ and $|\psi_2\rangle = -\sin\theta |e\rangle|0\rangle + \cos\theta |g\rangle|1\rangle$.

When the detuning between the cavity mode and the atomic resonance is large, such that $\Delta_c = \omega_a-\omega_c \gg g_c$, the mixing angle becomes $\theta \approx g_c/\Delta_c \ll 1$, and $|\psi_1\rangle$ is predominantly an atomic excitation with a small photonic component.
A second atom introduced into the cavity can then exchange an excitation with the first via the weakly populated cavity mode, leading to an effective interaction. For $N$ atoms this gives \cite{Agarwal1995}
\begin{equation}
H_I = \frac{\hbar g_c^2}{\Delta_c}\sum^N_{j,l}  \sigma_{eg}^j \sigma_{ge}^l,
\label{eq:cavity_int}
\end{equation}
which describes exchange of excitations between atoms with strength that does not diminish with distance, only being bounded by the volume of the physical cavity. These effectively infinite range interactions, while interesting in their own right \cite{Plenio1999a,Domokos2002a,Black2003a,Baumann2010a}, 
remove the spatial complexity of the system, and can often be described using collective operators or mean-field methods.

To realize long-range interactions that decay with distance we utilize photonic crystals. Key to our proposal is that through constructive interference of light scattering from the crystal's periodic structure, frequency windows known as band gaps can be created in which no propagating modes exist. Fig.~\ref{fig:cavity_prog}c shows a typical dispersion relation of photon frequency $\omega_k$ versus Bloch wavevector $k$ with a band gap.
Conventionally, a localized photonic crystal cavity mode is created by introducing a local dielectric defect (Fig.~\ref{fig:cavity_prog}b) that pulls a discrete mode into the band gap from the continuous band spectrum \cite{Joannopoulos2008}.
Here, we show that an atom trapped near the photonic crystal is itself a dielectric defect capable of seeding a cavity mode localized around the atomic position, via which it can interact with other atoms~(Fig.~\ref{fig:cavity_prog}d).

\section*{Atom induced cavities and long-range interactions}

The interaction between atoms and band edges has been discussed in a number of contexts, such as the formation of atom-photon bound states \cite{John1990a}, radiative coupling between atoms \cite{Kurizki1990a,John1991a,Bay1997a}, and spin-entanglement \cite{Shahmoon2013a} and thermalization \cite{Shahmoon2014a} mediated by long-range interactions.
Here, we provide an elegant interpretation of this physics in terms of atom-induced cavities and cavity QED. This mapping enables the powerful toolbox of cavity QED to be transferred to these systems, and enables one to identify key figures of merit (such as mode volume and cooperativity parameter). We exploit this mapping to demonstrate that the type of spin interaction and the spatial range can be manipulated dynamically, enabling tunable access to a wide range of long-range interacting models. We also identify the limits imposed by system imperfections (such as losses and disorder), and analyze in detail a realistic structure wherein this novel long-range physics can be realized.

A simple model to illustrate this mechanism consists of two-level atoms coupled to the photonic crystal modes, where the atomic resonance is close to one of the band edges of the photonic crystal (Fig.~\ref{fig:cavity_prog}c) with detuning $\Delta=\omega_a-\omega_b$. We assume that the detuning to any other band edge is much larger than $\Delta$ so that the influence of other bands is negligible.
When the atomic resonance is close to the band edge, the atom is dominantly coupled to modes close to the band edge wavevector $k_0$ due to the van Hove singularity in the density of states. In this case we can approximate the dispersion relation to be quadratic $\omega_k\approx\omega_b(1-\alpha(k-k_0)^2/k_0^2)$ about $k_0$, where $\alpha$ characterizes the band curvature \cite{John1990a}. Due to the periodicity of the photonic crystal the photonic modes are of Bloch form and the modes with wavevector $k\sim k_0$ take the form $E_k(z)\approx e^{i k z}u_{k_0}(z)$, with annihilation operator $\hat{a}_k$. Furthermore, for these modes, the coupling $g = d_{eg}\sqrt{\omega_b/(4 \pi\hbar\epsilon_0 A)}$ is approximately independent of $k$, where $A$ is the mode cross-sectional area \cite{Hung2013a}.

\begin{figure}
\centering
\includegraphics{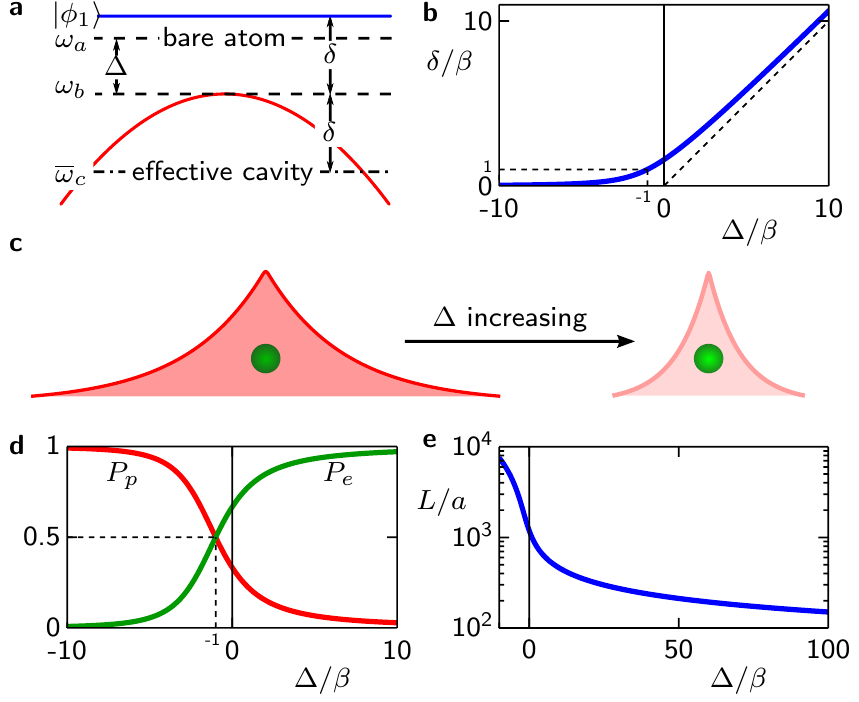}
\caption{\textbf{Effective cavity mode properties} (a) Energy level diagram for the photonic crystal dressed state $|\phi_1\rangle$ (blue).
The dressed state energy $\omega$ is detuned by $\delta$ from the band edge into the bandgap (band shown in red). The atom is coupled to an effective cavity mode with frequency $\overline{\omega}_c = \omega_b-\delta$ formed by superposition of modes in the band. (b) The detuning $\delta$ approaches 0 when $\Delta/\beta\ll -1$ and approaches $\Delta$ when $\Delta/\beta\gg 1$.
(c) The photonic component of the dressed state has an exponentially decaying envelope around the atomic position. Increasing $\Delta$ decreases the length scale $L$ of the exponential decay and the photonic part of the bound state superposition.
(d) The atomic excited state population of $|\phi_1\rangle$, $P_e = \cos^2(\theta)$ (green), increases as a function of $\Delta$, while the population of the photon mode, $P_p = \sin^2(\theta)$ (red), decreases as the state switches from photonic to atomic. (e) The length of the effective cavity decreases with $\Delta$. Here $L$ is in units of the lattice constant $a$, calculated for $\alpha = 10.6$ and $\beta = 4.75\times 10^{-7} \omega_b$, which is consistent with the ``alligator'' photonic crystal waveguide (see main text) \cite{Yu2013a}.}
\label{fig:cavity_vs_pc}
\end{figure}

A system with one atom trapped at $z=0$ coupled to the photonic crystal is then described by the Hamiltonian
\begin{equation}
H = \hbar \omega_a \sigma_{ee}+ \int dk \hbar \omega_k \hat{a}^\dagger_k \hat{a}_k+\hbar g\int dk (\sigma_{eg} \hat{a}_k E_k(0) +\mbox{h.c.}).
\label{eq:phc_ham}
\end{equation}
For a single excitation in the system, solving the Schr\"{o}dinger equation, $H |\psi\rangle = \hbar \omega |\psi\rangle$, yields the dressed state $|\phi_1\rangle = \cos\theta |e\rangle|0\rangle + \sin\theta |g\rangle|1\rangle$, where the atom is dressed by a localized photonic mode $|1\rangle = \int d k c_k \hat{a}_k^\dagger|0\rangle$  (see Supplementary Information (SI)).
The eigenfrequency $\omega$ lies within the band gap, with detuning $\delta=\omega-\omega_b>0$ from the band edge, as shown in Fig.~\ref{fig:cavity_vs_pc}a-b. The detuning $\delta$ is the positive real root of $(\delta-\Delta)\sqrt{\delta} = 2\beta^{3/2}$ for $\beta = \left(\pi g^2|u_{k_0}(0)|^2 k_0/\sqrt{4\alpha\omega_b}\right)^{2/3}$. 

Importantly, the photonic component is localized around the atomic position, as illustrated in Fig.~\ref{fig:cavity_vs_pc}c, with spatial mode function
\begin{equation}
\phi(z) =\int d k c_k^* E_k(z) = \sqrt{\frac{2\pi}{L}}e^{-|z|/L}E_{k_0}(z).\label{eq:phi}
\end{equation}
The photon decays exponentially with distance $z$ from the atomic position with length scale $L = \sqrt{\alpha\omega_b/( k_0^2\delta)}$, reflecting the fact that within the band gap at energy $\omega$, the field propagation equation has complex solutions with attenuation length $L$.


This confined photonic cloud has the same properties as a real cavity mode, enabling a mapping onto the Jaynes-Cummings model. Specifically, one can associate an effective atom-cavity interaction strength $\overline{g}_c = g\sqrt{2 \pi/L}$ with the bandgap system that depends on the mode volume,
expressed here as the effective cavity length given by the decay length $L$ for fixed mode area $A$. We can also identify an effective cavity mode frequency that is the average frequency of the photon mode, $\int d k |c_k|^2 \omega_k = \omega_b -\delta$. The effective atom-cavity detuning is then $\overline{\Delta}_c = \Delta+\delta$, as shown in Fig.~\ref{fig:cavity_vs_pc}a. In particular, for $\Delta = -\beta$ the effective cavity is resonant with the atom $\overline{\Delta}_c= 0$. The state $|\phi_1\rangle$ then maps to the dressed state $|\psi_1\rangle$ from the Jaynes-Cummings model, i.e., the mixing angle and energy are the same for $\overline{\Delta}_c\rightarrow\Delta_c$ and $\overline{g}_c\rightarrow g_c$. 
This mapping breaks down when we consider the second dressed state, which is not an eigenstate in the photonic crystal model because of the continuum of propagating modes for frequencies below $\omega_b$.

The atomic excited state population $P_e=\cos^2\theta$ for $|\phi_1\rangle$ is plotted in Fig.~\ref{fig:cavity_vs_pc}d. For $\Delta\gg \beta$, most of the excitation resides in the atom, while for $\Delta \ll -\beta$ the state becomes mostly photonic, a cavity mode dressed by the atom. Physically, although the atomic frequency can lie well within the band, the atom still provides a weak refractive index contrast at frequencies within the gap, and in 1D, arbitrarily weak dielectric defects can seed a cavity mode \cite{MarkosWP}. In this regime (Fig.~\ref{fig:cavity_vs_pc}e), the weak dielectric contrast yields a very long effective cavity length. In practice, this length is only limited by the finite size of the photonic crystal structure or by disorder in the photonic crystal structure~(as discussed below).

It is now apparent that two atoms can exchange an excitation via the induced cavity mode provided they are separated by a distance of order of the decay length $L$. In the limit where the photonic modes are weakly populated ($\Delta\gg\beta$), this leads to an effective dipole-dipole interaction between atoms, with positions $z_j$, of the form (see SI) 
\begin{equation}
H_I \approx \frac{\hbar \overline{g}_c^2}{\overline{\Delta}_c}\sum^N_{j,l}\sigma_{eg}^j\sigma_{ge}^l f(z_j,z_l).
\label{eq:dipole_int}
\end{equation}
 The effective atom-cavity detuning is $\overline{\Delta}_c=2\Delta$ (since $\delta\sim\Delta$) and $f(z_j,z_l) = e^{-|z_j-z_l|/L}E_{k_0}(z_j)E^*_{k_0}(z_l)$ \cite{Shahmoon2013a}. 

Compared to the case of a conventional cavity
(Eq.~(\ref{eq:cavity_int})), the main feature of interactions emerging from atom-induced cavities is that the spatial function $f(z_j,z_l)$ is finite-range and tunable, both through the effective interaction length $L$ and the Bloch functions~(e.g., from which changes in sign can be engineered).  In addition, this dynamic cavity mode follows the atomic position rather than being a static property set by boundary conditions. Note, that although exponentially decaying interactions are identified as short-range in the thermodynamic limit, the length scale $L$ can be of the order of the length of experimental systems, e.g., effectively long-range over the system size \cite{Shahmoon2013a,Shahmoon2014a}. Furthermore, as we show below, the interaction can approximate long-range power-laws over a finite system, similar to in trapped ion experiments \cite{Porras2004a,Islam2013a,Richerme2014a,Jurcevic2014a}. These results also generalize to higher dimensions (see SI), e.g., in a two-dimesional photonic crystal atom-induced cavities lead to a function $f(\mathbf{z}_j,\mathbf{z}_l)\approx E_{\mathbf{k}_0}(\mathbf{z}_j)E^*_{\mathbf{k}_0}(\mathbf{z}_l)e^{-|\mathbf{z}_j-\mathbf{z}_l|/L}/\sqrt{|\mathbf{z}_j-\mathbf{z}_l|}$ \cite{Tudela2014a}.

While Eq.~(\ref{eq:dipole_int}) superficially looks like a long-range spin model, the possible dynamics are in fact much richer. In particular, treating the atomic positions themselves as dynamical variables, the function $f(z_j,z_l)$ can physically be interpreted as a mechanical potential acting on atoms \cite{Shahmoon2014a}, which is turning on and off as spin degrees of freedom change. Due to the large values of the effective vacuum Rabi splittings $\overline{g}_c$ associated with the dynamic atomic cavities, the strength of these spin-dependent potentials can be extremely large compared to typical motional energy scales associated with ultracold atoms. Furthermore, in the nanophotonic system, it is possible to achieve strong spin-photon coupling (for example, an incident photon can be absorbed by the atoms with high probability \cite{Goban2013a}). Thus, our system produces a unique coupling over long range between spin, phonon, and photon degrees of freedom.

\section*{Coherence and effective cooperativity}

Key to any physical realization of the long-range physics we describe above, is how dissipation competes with the coherent interaction in Eq.~\ref{eq:dipole_int}. Here, we go beyond previous discussions of this type of interaction, by detailing the limits imposed by realistic loss mechanisms and experimental imperfections, and further show these interactions should be observable in current state of the art experiments.

Imperfections in the photonic crystal cause photon loss at rate $\kappa_{p}$, and because our structures of interest are not full 3D photonic crystals, an excited atom can spontaneously emit into free space at rate $\gamma$~(usually comparable to the vacuum rate $\gamma_0$ \cite{Hung2013a}). The effect of losses can be revealed, for example, by studying the exchange of
an excitation between two atoms separated by $|z_1-z_2|\lesssim L$. From Eq.~(\ref{eq:dipole_int}), the exchange time is given by $\tau \sim \pi\overline{\Delta}_c/(2\overline{g}_c^2)$, while the total loss is given by $\tau (\gamma \cos^2 \theta + \kappa_{p} \sin^2 \theta)$. Optimizing the detuning, we find an exchange error of $\pi/\sqrt{C}$, where $C =\overline{g}_c^2/(\kappa_{p}\gamma)$ is the single-atom cooperativity  (see Methods). For a state of the art photonic crystal ($Q \sim 200000$) coupled to Cesium atoms ($\gamma/(2 \pi)\sim  5$MHz), a cavity with volume $V\sim\lambda^3$ (i.e.~$L=\lambda$) could have $\overline{g}_c/(2 \pi) \sim 10$GHz, giving a feasible cooperativity of $C_\lambda \sim 10^4$ \cite{Yu2013a}. Assuming that $\kappa_{p}$ is dominated by local imperfections in the photonic crystal, we expect that it is independent of the length $L$~(as compared to a Fabry-Perot cavity in vacuum, where $\kappa_p\propto 1/L$). The cooperativity of the dynamic cavity mode then scales with length as $C_L = \lambda C_\lambda /L$, which limits the length for which interactions remain coherent. For $C_\lambda \sim 10^4$, $C_L$ remains greater than 100 for lengths of up to 100 wavelengths.

Beyond the photon losses already discussed, imperfections in photonic crystal fabrication can yield disordered potentials for propagating fields. Disorder results in an Anderson localization length over which optical fields tend to become trapped, limiting the interaction range between atoms. As discussed in the Supplementary Information, the physics of weak disorder near a band edge leads to a universal scaling for the localization length as a function of the level of disorder. This scaling predicts that localization lengths exceeding 100 wavelengths are possible with state-of-the-art fabrication and is consistent with separate (unpublished) experimental characterization of the ``alligator'' structure from Ref.~\cite{Yu2013a}, which reveals a localization length longer than the structure length of $\sim 200$ unit cells. On the other hand, disorder may prove to be a feature of the system, providing access to physical models with interactions that have randomly varying length scales.

\section*{Implementation in an ``alligator'' photonic crystal waveguide}


\begin{figure}
\centering
\includegraphics{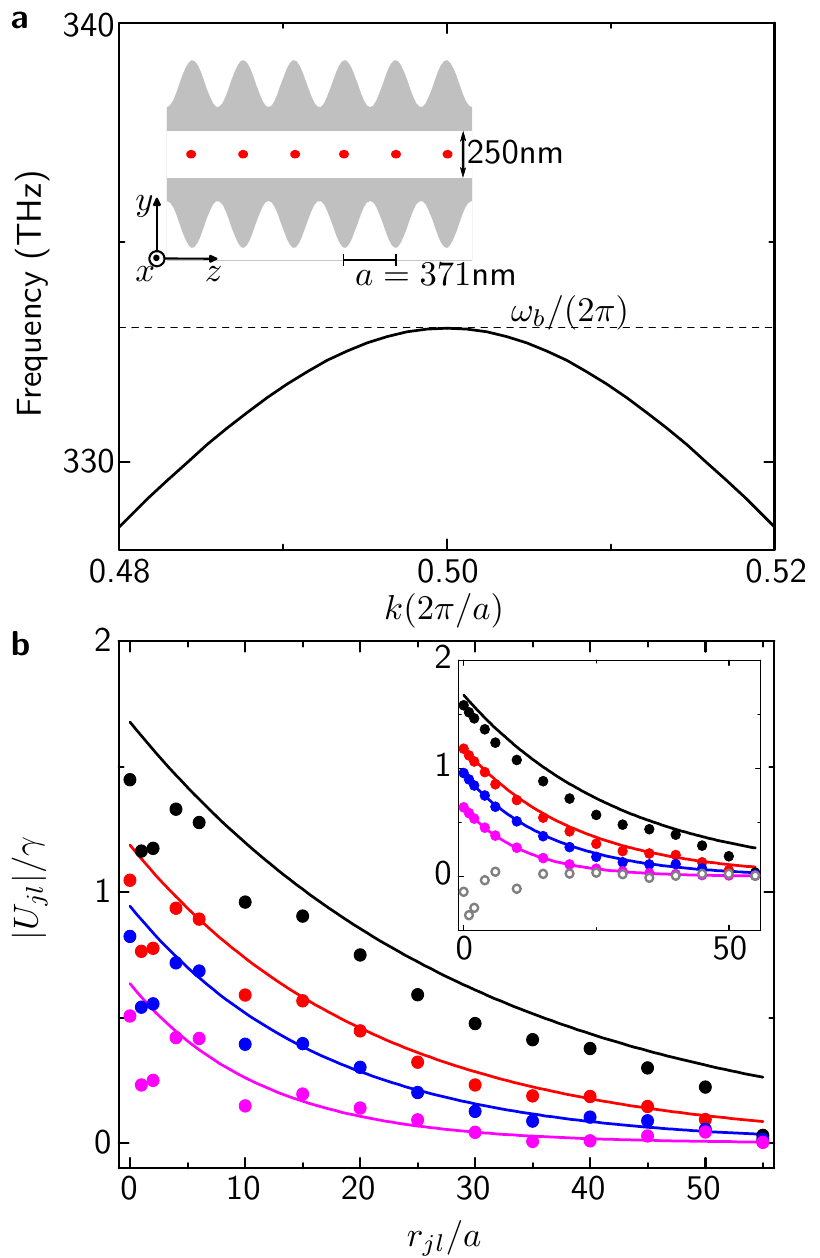}
\caption{\textbf{Comparison of the single-band model with numerical calculations.} (a) Band structure of fundamental transverse-electric (TE) mode of the sample 1D ``alligator'' photonic crystal waveguide (APCW), designed for coupling to the D1 line of atomic cesium near the photonic bandedge frequency $\omega_b/2\pi=333~$THz \cite{Goban2013a,Yu2013a}. The calculated band structure has a curvature $\alpha \approx 10.6$ near the band edge at $k_0 = \pi/a$. Inset shows the dielectric profile of the APCW. Red circles denote the location of trapped atoms. (b) Atom-atom coupling strength $U_{ij}$ evaluated using finite-difference time-domain simulations (solid circles) and the single-band model from Eq.~(\ref{eq:dipole_int}) (solid lines). Results are plotted for atomic detunings from the band edge $\Delta/2\pi =$ 400 (black), 800 (red), 1300 (blue), 2800 (magenta) GHz.  Inset shows the FDTD results where 
contribution from all other photonic and free-space modes in the APCW (open circles) has been estimated numerically and subtracted; see Methods.}\label{fig:alligator}
\end{figure}


Our simple theoretical model should approximate well actual physical implementations. As a concrete example, we consider the one-dimensional ``alligator'' photonic crystal waveguide~(APCW)  as experimentally demonstrated in Ref.~\cite{Goban2013a,Yu2013a}. The APCW, as shown in Fig.~\ref{fig:alligator}a, consists of two parallel, periodically corrugated nanobeams, whose small distance of separation couples and hybridizes their optical modes. A combination of far off-resonant guided modes and Casimir-Polder forces allows atoms to be localized between the beams at the periodic points indicated in red in Fig.~\ref{fig:alligator}a \cite{Hung2013a}. 

The band structure of the APCW, calculated using the MIT Photonic-Bands software package, is shown in Fig.~\ref{fig:alligator}a. The band edge of the fundamental transverse electric (TE-like) mode, located at $\omega_b/2\pi=333$~THz, is closely aligned with the D${}_1$ transition of atomic cesium to produce the desired long-range interactions. In Fig.~\ref{fig:alligator}b, we plot the coefficients $|U_{jl}|/\gamma$ (see Methods) for the effective atom-atom interaction $H_I=\hbar\sum_{j,l}U_{jl}\sigma_{eg}^{j}\sigma_{ge}^l$ associated with the APCW. Couplings are plotted for detunings of the atom from the band edge ranging from $\Delta/2\pi=400$~GHz to $2.8$~THz, over atomic separations $r_{jl}/a$ extending up to $55$ lattice sites~($a=371$~nm is the lattice constant of the APCW). 
The predictions from our simple model, Eq.~(4)~(solid lines in Fig.~\ref{fig:alligator}b), quantitatively agree with the numerical simulation of the full structure for the structure band curvature $\alpha=10.6$ and coupling $\bar{g}_c /2\pi=\sqrt{a/L}\times 12.2$~GHz (see Methods). The deviation between the theoretical and numerical results at $\Delta/2\pi=400$~GHz is primarily attributable to finite-size effects, as the interaction length becomes comparable to the simulated structure size~(75$a$ between the source and either end of the APCW). 

At short atomic separations $r_{jl}/a\lesssim 15$, numerical results deviate from the model at all detunings (Fig.~\ref{fig:alligator}b).
We primarily attribute this difference to the contributions coming from other guided bands (see Ref.~\cite{Goban2013a} for full band structure), as well as leaky and free-space modes that are not included in our single-band theoretical model.  The band edges of these modes are typically far from the atomic transition frequency, leading to contributions of the order of the interaction strength between atoms in free space or coupled to a nanofibre, that is, at most the free-space linewidth $\gamma$ \cite{GH82,Goban2012a}. The fractional error associated with these corrections will then become smaller as experimental systems are optimized to increase the interaction strength arising from the primary mode. This can be done by working closer to the band-edge (seen to only a limited extent in Fig.~\ref{fig:alligator}b because of finite size effects), or by using structures with smaller band curvature $\alpha$ to decrease the interaction length.
Even better agreement with the model can be reached by approximately subtracting out the contributions from the other modes (see Methods), as plotted in the inset of Fig.~\ref{fig:alligator}b, supporting the validity of our simple model.

\section*{Designing interaction properties}

\begin{figure}
\centering
\includegraphics{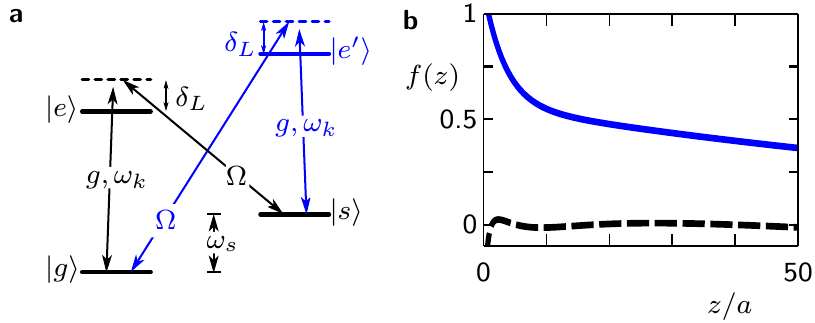}
\caption{\textbf{Designing interaction potentials} (a) Driven (black) $\Lambda$ and (black and blue) four level system. In the $\Lambda$ scheme, transition $|g\rangle$-$|e\rangle$ couples with strength $g$ to the photonic crystal modes, while $|s\rangle$-$|e\rangle$ is pumped by a laser with detuning $\delta_L$ and Rabi frequency $\Omega$. Interactions between the $x$-component of the effective spin can be achieved by adding level $|e'\rangle$, where the transition $|s\rangle$-$|e'\rangle$ also couples to the modes of the photonic crystal, while a second pump
drives $|g\rangle$-$|e'\rangle$.
(b)  Approximate power law interactions between atoms over a finite region can be achieved by summing the different exponential interactions associated with multiple drive fields. This is illustrated here over 50 lattice sites, where two exponentials are added to yield a $\eta=1/4$ power law, $f(z)=w_1 e^{-s_1 z/a}+w_2 e^{-s_2 z/a} \approx z^{-1/4}$ (solid blue curve). The error $f(z)-z^{-1/4}$ is given by the dashed curve. Here $w_1 = 0.5480, w_2 = 0.5684, s_1 =  0.2916$ and $s_2 =  0.0089$ could be achieved by detuning one laser from the band edge by $1.723\times 10^{-3}\omega_b$ and the second by $1.612\times 10^{-6}\omega_b$ for $\alpha=0.2$.}\label{fig:localization}
\end{figure}

The long-range interactions given in Eq.~(\ref{eq:dipole_int}) depend on the detuning from the band edge $\Delta$ and band curvature $\alpha$, which cannot be easily tuned given a physical structure. This is remedied by considering atoms with an internal $\Lambda$-level structure, as shown in Fig.~\ref{fig:localization}a, introducing an additional metastable state $|s\rangle$. Here the $|g\rangle$-$|e\rangle$ transition is coupled to the band edge as before, while $|s\rangle$-$|e\rangle$ is assumed to be de-coupled but addressable by an external laser (e.g.~illuminating the photonic crystal from the side) with Rabi amplitude $\Omega$ and detuning $\delta_L$. This situation may be achieved, for example, if the photonic crystal modes and external laser have orthogonal polarizations. Alternatively, guided modes of the photonic crystal with orthogonal polarization may be used.

The photons in the system are now Raman scattered from the laser field with central frequency $\omega_a+\delta_L$, and intuitively the photon cloud size will be determined by the attenuation length at this frequency rather than at $\omega_a$ as in the case of pure atomic excitation. Specifically, adiabatically eliminating the excited state and the photonic modes (see SI), we obtain an interaction within the ground-state manifold of the form in Eq.~(\ref{eq:dipole_int}),
\begin{equation}
H_I = \frac{\hbar|\Omega|^2 \overline{g}_c^2}{2\Delta_L \delta_{L}^2}\sum^N_{j,l}\sigma_{sg}^j\sigma_{gs}^l f(z_j,z_l),\label{eq:XY_mod}
\end{equation}
where now $\Delta = \omega_a-\omega_b$ is replaced by $\Delta_L = \delta_L+\omega_a-\omega_b$ and $\overline{g}_c$ is replaced by  $\Omega\overline{g}_c/\delta_{L}$.
The strength of the interaction is reduced by a factor of $|\Omega|^2/\delta_{L}^2$, leading to an increase in the time needed for the spin exchange; however, the population of the atomic excited state is also reduced by the same factor reducing the rate of spontaneous emission. The Raman process effectively narrows the natural line width of the excited state, and the cooperativity $C$, which characterizes the optimal fidelity of exchange, remains constant.

By tuning the frequency and amplitude of the drive laser, the interaction strength and length  $L = \sqrt{\alpha\omega_b/(\Delta_L k_0^2)}$ can now be dynamically altered. It also becomes feasible to build interaction scalings other than exponential, by driving the $|s\rangle$-$|e\rangle$ transition with two or more fields of different frequencies. This leads to an interaction potential for the atoms which is the sum of the potentials due to each individual drive field (the adiabatic elimination of drive fields is additive).
For example, a power law interaction $f(z_j,z_l)\propto|z_j-z_l|^{-\eta}$ can be approximated over a finite range, as we show in Fig.~\ref{fig:localization}b for $\eta=1/4$ over 50 unit cells using two pump fields.


The full spin-phonon-photon dynamics resulting from the coupling of atoms to the photonic crystal bands enables a wide range of behaviour to be investigated. In addition, the limiting cases of the model, found by freezing out degrees of freedom, are by themselves interesting. For example, the motional modes may be eliminated by trapping atoms tightly on a lattice to produce quantum magnetism models, such as the XY-model and the transverse Ising model (see Methods), where
long-range interactions lead to a breakdown of Lieb-Robinson bounds \cite{Hauke2013a,Richerme2014a,Jurcevic2014a}. Alternatively, in the opposite limiting case, we may focus on the motional dynamics of the atoms, eliminating the spin dynamics by driving the atoms weakly and off resonance. In this case the interaction yields a purely mechanical potential for the atoms,
$U\approx \frac{\hbar|\Omega|^2 g_c^2}{2(\omega_L-\omega_b)(\omega_L-\omega_a)^2} f(z_j,z_l)$.
Engineering the interaction to be a power law with $\eta=1$, for example, would enable simulation of charged particles using neutral cold atoms.

\section*{Discussion}

In conclusion, we have demonstrated the utility of atoms coupled to photonic crystal structures as a toolbox to achieve tunable long-range interactions in many-body systems. Moreover, significant progress has been made to experimentally realize systems where the predicted rich interplay between spin, motional and photonic degrees of freedom could be observed. Individual atoms have been trapped and coupled to a photonic crystal cavity using optical tweezers \cite{Thompson2013a}, and atoms may also be trapped in the evanescent field of guided modes of photonic crystals \cite{Hung2013a,Goban2013a,Yu2013a}.

In these systems, many-body phenomena, such as frustration \cite{Mattioli2013a}, information propagation \cite{Hauke2013a,Richerme2014a,Jurcevic2014a} and many-body localization \cite{Basko2006a} may be investigated in the presence of long-range interactions. Such interactions could also
generate non-local, non-linearities for photons, leading to photonic molecules and other exotic states \cite{Longo2011a,Firstenberg2013a}.
Finally, the use of band edges to manipulate interactions should find wide use beyond the photonic crystal setting. In the context of atom-photon interactions, band structure could be engineered using a variety of fabrication techniques \cite{Nayak2013a,Yu2013a}, 
or using periodic arrangements of atoms themselves \cite{Chang2012a}, while 
coupling of quantum bits via phononic band structure \cite{Eichenfield2009a} could be controlled in an analogous manner.

\section*{Methods}

\subsection*{Coherence and effective cooperativity}

The optimal loss rate connected with the cooperativity parameter is reached by adjusting the detuning of the effective cavity. However, adjusting only the detuning also changes the length scale. To keep the length $L = \sqrt{\alpha\omega_b/(\Delta k_0^2)}$ fixed and achieve the optimal detuning we must also adjust the band curvature $\alpha$.  
The APCW photonic crystal structures to which atoms have already been successfully coupled have curvature parameters on the order of $\alpha\sim 10$, giving an optimal length of 
$L\sim 100 \lambda$ with $C_L\sim 100$ for the parameters above. Slight design changes should enable values of $\alpha\sim 1$  leading to optimal length $L\sim 20\lambda$ and 
$C_L\sim 600$ \cite{Hung2013a}. Reaching shorter interaction lengths with a fidelity limited by $C_L$ requires structures with even flatter bands, and the design of such systems that are also compatible with atom trapping is being investigated further. 

\subsection*{Implementation in an ``alligator'' photonic crystal waveguide}

The couplings $U_{jl}$ are calculated assuming the atoms are at the minima of the APCW trapping potentials and the atomic transition is polarized along $\hat{y}$~(the local direction of polarization for the fundamental TE mode). The $U_{jl}$ are directly proportional to the real part of the dyadic electromagnetic Green's function $G_{yy}(r_j,r_l,\omega_a)$~\cite{Dung2002a}. The Green's function physically describes the field produced at position $r_j$ due to a source at $r_l$ and frequency $\omega_a$, and is obtained for the APCW by finite-difference time-domain~(FDTD) simulations.
The coupling coefficients normalized by the free-space emission rate are $|U_{jl}|/\gamma_0=|\textrm{Re}\,G_{yy}(r_j,r_l,\omega_a)|/(2\,\textrm{Im}\,G_{yy,\footnotesize\textrm{free}}(0,0,\omega_a))$, where $G_{\footnotesize\textrm{free}}$ is the free-space Green's function. To check the validity of our theoretical model given by Eq.~(\ref{eq:dipole_int}), we use a parameter of $\alpha=10.6$ describing the actual curvature near the band edge for the APCW \cite{Yu2013a}. The atom-field coupling strength $\bar{g}_c$ is obtained from first principles by numerical quantization of the guided modes near the band edge~(see SI), yielding $\bar{g}_c /2\pi=\sqrt{a/L}\times 12.2$~GHz.

To estimate the contribution to the interaction coefficient from other modes, we note that leaky and free-space modes should exhibit negligible frequency dependence over the narrow range of detunings $\Delta$ considered. We then estimate the multimode contribution by calculating the difference between the model and numerical $|U_{jl}/\gamma_0|$ at $\Delta/2\pi = 3$THz.

\subsection*{Designing interaction properties}

The interaction in Eq.~(\ref{eq:XY_mod}) corresponds to the XY-model, $\sum_{j\neq l}(\sigma_{x}^j
\sigma_{x}^l +\sigma_{y}^j
\sigma_{y}^l)f(z_j,z_l)$, for spin-1/2 operators $(\sigma_x,\sigma_y,\sigma_z) = (\sigma_{gs}+\sigma_{sg},i(\sigma_{gs}-\sigma_{sg}),\sigma_{ss}-\sigma_{gg})/2$. Other level schemes, such as the four level structure in Fig.~\ref{fig:localization}a, enable other spin models. Here, the $\Lambda$-structure is extended by adding a transition $|s\rangle$-$|e'\rangle$ that also couples to the photonic crystal modes, while a second pump with the same amplitude and detuning as the first drives the $|g\rangle$-$|e'\rangle$ transition. Eliminating the excited state manifold (see SI) now yields an effective Hamiltonian corresponding to the transverse Ising model with long-range interactions,
\begin{equation}
H = \hbar\omega_s\sum^N_j \sigma_z^j + \frac{2\hbar|\Omega|^2 \overline{g}_c^2}{\Delta_L\delta_{L}^2}\sum^N_{j\neq l}\sigma_{x}^j
\sigma_{x}^l f(z_j,z_l).
\label{eq:XX_mod}
\end{equation}

\section*{Acknowledgments}
The authors thank L.~Tagliacozzo, P.~Hauke, M.~Lewenstein, A.~Gonz\'alez-Tudela, J.~I.~Cirac, L.~Jiang, J.~Preskill, O.~Painter, M.~Lukin, J.~Thompson, and S.~Gopalakrishnan for insightful discussions.
This work was supported by Fundacio Privada Cellex Barcelona, the MINECO Ramon y Cajal Program, the Marie Curie Career Integration Grant, the IQIM, an NSF Physics Frontiers Center, the DoD NSSEFF program, DARPA ORCHID, AFOSR QuMPASS MURI, NSF
PHY-1205729, NSF PFC at the JQI, NSF PIF, ARO, ARL, and AFOSR MURI on Ultracold Polar Molecules.


\pagebreak
\onecolumngrid
\renewcommand{\thefigure}{S\arabic{figure}}
\renewcommand{\theequation}{S\arabic{equation}}
\section*{Supplementary information: Realizing quantum many-body models with cold atoms coupled to photonic crystals}

\section{Single excitation bound state}

For a two-level atom interacting with the electromagnetic field modes of a photonic crystal, a bound state exists that is the superposition of an atomic excitation and a localized excitation of the photonic modes \cite{John1990a,John1991a,Kofman1994a,John1994a}.
The atom-photon bound state is the solution to the Schr\"{o}dinger equation, $H |\psi\rangle = \hbar \omega |\psi\rangle$, where $\omega$ lies within the band gap, for the Hamiltonian given in Eq.~(2) of the main text. We assume the bound state has frequency near the lower band edge with detuning $\delta=\omega-\omega_b>0$. The upper band is assumed to be much further detuned from the atomic resonance and have a negligible effect on the bound state (the complementary solution near the upper band is found in the same manner).

Making the effective mass and associated approximations described in the main text, the eigenenergy is found to be the positive real root of $(\delta-\Delta)\sqrt{\delta} = 2\beta^{3/2}$ for $\beta = \left(\pi g^2 |u_{k_0}(0)|^2k_0/\sqrt{4\alpha\omega_b}\right)^{2/3}$ and $\Delta = \omega_a-\omega_b$.
The solution can be expressed explicitly as
\begin{equation}
\delta = \frac{2}{3}\Delta+\beta\left(\lambda_+^{1/3}+\lambda_-^{1/3}\right)
\quad \mbox{where} \quad
\lambda_{\pm} = \left(1\pm\sqrt{1-\frac{\Delta^3}{27\beta^3}}\right)^2.
\end{equation}
The corresponding eigenstate is expressible in the form $|\phi_1\rangle = \cos\theta |e\rangle + \sin\theta |1\rangle$ where
$|1\rangle = \int d k c_k^* \hat{a}_k^\dagger|0\rangle$ is a single photon excitation of the modes in the band. Here
$\cos\theta = (1+(\beta/\delta)^{3/2})^{-1/2}$, $\sin\theta = (1+(\delta/\beta)^{3/2})^{-1/2}$ and $c_k = (\delta/\beta)^{3/4}g u_{k_0}(0)/(\omega-\omega_k)$. In the limit $\Delta \gg\beta$, the detuning $\delta \rightarrow \Delta$ and the population of the excited state $\cos^2\theta\rightarrow 1$.

The weights of the photonic modes give us an indication of the validity of the approximations used above to derive the form of the bound state. The distribution $c_k \propto (\delta+\alpha\omega_b(k-k_0)^2/k_0^2)^{-1}$ is a Lorentzian in $k$ centered at $k_0$ with half width $\sqrt{\delta/(\alpha\omega_b)}k_0$. Our approximations required that the populated modes have wavevectors close to $k_0$, that is, provided $\sqrt{\delta/(\alpha\omega_b)}\ll 1$.

\section{Dipole-dipole interactions}

\subsection{Two-level atoms}

We first consider the case of many two-level atoms trapped near a photonic crystal. The atomic resonance frequency is taken to be in the band gap near the lower band edge, such that the detuning satisfies $\Delta\gg\beta$. In this case the photonic modes are weakly populated and we may eliminate them to obtain a description of the system in terms of effective dipole-dipole interactions \cite{Kurizki1990a,John1991a,Bay1997a,Shahmoon2013a} 
as detailed below.

In the interaction picture, the interaction between the atom and photonic crystal modes is
\begin{equation}\label{eq:inter_H}
\mathcal{H}_I = \hbar \sum_j\int dk g_k\sigma^j_{eg} \hat{a}_k u_k(z_j) e^{i \delta_k t+i k z_j} +\mbox{H.c.},
\end{equation}
where $\delta_k=(\omega_a-\omega_k)$.
In the presence of loss, we can describe the evolution of the system  by the master equation
$\dot{\rho}=\mathcal{L}_{int}(t)\rho+\mathcal{L}_\gamma\rho+\mathcal{L}_\kappa\rho$.
Here
$\mathcal{L}_{int}(t)\rho = -\frac{i}{\hbar}[\mathcal{H}_{I},\rho]$
describes the coherent evolution, while the loss mechanisms resulting from spontaneous emission from the atomic excited state (rate $\gamma$) and loss of photons in the photonic crystal (rate $\kappa$) are described by
\begin{equation}
\mathcal{L}_\gamma\rho=-\frac{\gamma}{2}\sum_j \left( \{\sigma_{ee}^j,\rho\}-2\sigma_{ge}^j\rho\sigma_{eg}^j \right) \quad \mbox{and} \quad
\mathcal{L}_\kappa\rho=-\frac{\kappa}{2}\int dk \left( \{\hat{a}^\dag_k \hat{a}_k,\rho\}-2\hat{a}_k\rho \hat{a}^\dag_k \right).
\end{equation}
We now proceed to eliminate the photonic modes using the Nakajima-Zwanzig approach in the Born-Markov approximation \cite{Breuer02}. This yields the master equation for the reduced density operator, $\rho_s = \rm{Tr}_k(\rho)$ after tracing over the field modes,
\begin{equation}
\dot{\rho}_s=-\frac{i}{\hbar}[\mathcal{H}^{ef}_I,\rho]+\mathcal{L}_\gamma\rho_s+\mathcal{L}^{ef}_\kappa\rho_s.
\end{equation}
With the field modes eliminated, the interaction is described by the effective Hamiltonian
\begin{equation}\label{eq:two_lev_modes}
\widetilde{\mathcal{H}}^{ef}_I = \sum_{j,l}\sigma^j_{eg}\sigma^l_{ge}\int dk \frac{g_k^2 \delta_k E_{k}(z_j)E^*_{k}(z_l)}{\delta_k^2 +(\kappa/2)^2}.
\end{equation}

We can then integrate over the field modes, applying the same approximations near the band edge as used to derive the bound state, that is, the band is quadratic and only makes significant contributions for $k\sim k_0$. To first order in the $\kappa/\Delta$, the interaction becomes
\begin{equation}\label{eq:two_lev_ef}
{\mathcal{H}}^{ef}_I =  \frac{g_c^2}{2\Delta}\sum_{j,l}E_{k_0}(z_j)E^*_{k_0}(z_l)\exp[-|z_j-z_l|/L]\sigma^j_{eg}\sigma^l_{ge}.
\end{equation}
where we have defined $L = \sqrt{\alpha\omega_b/(\Delta k_0^2)}$ and $g_c=\sqrt{2 \pi/L}g$.
If we had considered atoms with resonance near the upper band instead, with $\alpha<0$, the above relation would still hold, however $\Delta$ would now be negative in the band gap and the interaction would hence have opposite sign. This provides a method to tune the sign of the interactions by placing the atomic resonance close to either the lower or upper band edges. To lowest order, the loss of coherence due to photon loss in the crystal is described by
\begin{equation}
{\mathcal{L}}^{ef}_\kappa\rho=-\frac{g_c^2\kappa}{8\Delta^2}\sum_{j,l}E_{k_0}(z_j)E^*_{k_0}(z_l)
\left(\{\sigma^j_{eg}\sigma^l_{ge},\rho\}-2\sigma^j_{ge}\rho\sigma^l_{eg} \right).
\end{equation}
The leading order loss terms are hence a factor of $\kappa/(4\Delta)$ smaller than the coherent interaction.

The interaction in Eq.~(\ref{eq:two_lev_ef}) is obtained by the integration in Eq.~(\ref{eq:two_lev_modes}) over a one dimensional band structure. For two or three dimensional band structure Eq.~(\ref{eq:two_lev_modes}) is generalized to become
\begin{equation}\label{eq:dipole_int_gen}
\mathcal{H}^{ef}_I  = \hbar g^2\sum_{j,l}\sigma_{eg}^j\sigma_{ge}^l \int d \mathbf{k}\frac{E^*_\mathbf{k}(\mathbf{z}_l)E_\mathbf{k}(\mathbf{z}_j)}{\omega_a-\omega_\mathbf{k}}.
\end{equation}
For example, for a two-dimensional band edge of the form $\omega_\mathbf{k} = \omega_b(1 - \alpha|\mathbf{k}-\mathbf{k}_0|^2/k_0^2)$, the interaction has spatial dependence $f(\mathbf{z}_j,\mathbf{z}_l)=
\frac{2}{\pi}
\mathcal{K}_0(|\mathbf{z}_j-\mathbf{z}_l|/L)E_{\mathbf{k}_0}(\mathbf{z}_j)E^*_{\mathbf{k}_0}(\mathbf{z}_l)$,
where $\mathcal{K}_0(z)$ is the Bessel function of the second kind.
The interaction now decays approximately exponentially with an additional $1/\sqrt{|\mathbf{z}_j-\mathbf{z}_l|}$ scaling due to the two-dimensional spread of the photon cloud.

\subsection{Weakly driven multi-level atoms}

We can now examine the effective interactions between atoms with more complicated internal level structures as depicted in Fig.~4a of the main text. We assume the transitions not coupled to the photonic crystal are weakly excited by an orthogonally polarized laser and we denote by $\Omega$ the Rabi driving of $|s\rangle$-$|e\rangle$ and by $\Omega'$ the drive of $|g\rangle$-$|e'\rangle$, which is zero in the three level case. For concreteness, we assume spontaneous emission occurs with equal rate $\gamma$ from both excited states $|e'\rangle$ and $|e\rangle$, and decay from each state occurs with equal probability into either of the ground states. We take the driving frequency to be detuned by $\delta_L\gg \gamma,\Omega^{(\prime)}$ from each atomic resonance, in which case the atoms are weakly driven and the Raman scattered fields are predominantly centered around the two-photon resonance frequency. When the laser frequency is in the photonic band gap we then expect the atoms to form bound states with photons near the laser frequency.

In this regime we can obtain effective dynamics for the long-lived atomic ground states $|s\rangle$ and $|g\rangle$. We follow a similar procedure as above. In particular, in a frame rotating at the laser frequency, we first eliminate the photonic modes using the Nakajima-Zwanzig technique, followed by the atomic excited states.
Following this recipe we obtain effective dynamics to first order in $\kappa/\Delta$ and $\gamma/\delta_L$ described by
\begin{equation}\label{eq:four_level_H}
\mathcal{H}^{ef}_{I} =  \frac{g_c^2}{2\Delta_L}\frac{|\Omega|^2}{\delta_L^2}\sum_{j,l}E_{k_0}(z_j)E^*_{k_0}(z_l)\exp[-|z_j-z_l|/L]
S_j^\dagger S_l.
\end{equation}
where $S_j= ((\Omega'/\Omega)\sigma_{sg}^j+\sigma_{gs}^j)$ and we make the replacement $\Delta\rightarrow\Delta_L  = \omega_L+\omega_s-\omega_b$ in the expression for $L$. The replacement of $\Delta$ by $\Delta_L$ means $L$ now changes with the laser frequency, leading to the possibility of dynamic tuning of the interaction length.

The loss of coherence in the ground state manifold resulting from spontaneous emission into free space is described by
\begin{equation}
\mathcal{L}_\gamma^{ef}\rho=-\sum_j\left[\frac{\tilde\gamma}{2}\left(\{\sigma_{ss}^j,\rho\}
-\sigma_{ss}^j\rho\sigma_{ss}^j-\sigma_{gs}^j\rho\sigma_{sg}^j \right)
+\frac{\tilde\gamma'}{2}\left(\{\sigma_{gg}^j,\rho\}-\sigma_{gg}^j\rho\sigma_{gg}^j -\sigma_{sg}^j\rho\sigma_{gs}^j \right)\right]
\end{equation}
where the effect of the Raman transitions is to narrow the linewidths of the transitions according to
$\tilde\gamma^{(\prime)}=|\Omega^{(\prime)}|^2\gamma/\delta_L^2$.
Losses in the photonic crystal induce another dissipation mechanism for the atomic degrees of freedom, which is given by
\begin{equation}
\mathcal{L}^{ef}_\kappa\rho=-\frac{g_c^2\kappa}{8\Delta_L}\frac{|\Omega|^2}{\delta_L^2}\sum_{j,l}E_{k_0}(z_j)E^*_{k_0}(z_l)
\left( \{S_j^\dagger S_l,\rho\}-2S_j\rho S_l^\dagger \right).
\end{equation}
The interaction and losses are all reduced by the same factor, $|\Omega|^2/\delta_L^2$, and as a result all processes happen on a slower time scale, while the cooperativity of the system remains constant.

From Eq.~(\ref{eq:four_level_H}) we obtain the effective interactions given in the main text in Eqs.~(5)-(6). 
We obtain the $\Lambda$-system interaction, Eq.~(5), when $\Omega'=0$ such that $S_j = \sigma_{gs}$, and the four level system interaction, Eq.~(6), when $\Omega' = \Omega$ such that $S_j = 2\sigma_x$.
For $\Omega' = \Omega e^{i\phi}$, the interaction in Eq.~(\ref{eq:four_level_H}) is rotated to be between spins $S_j= 2\cos(\phi/2)\sigma_{x}^j-2\sin(\phi/2)\sigma_{y}^j$. We can then adjust the spin-spin interaction temporally or spatially by adjusting $\phi$.

\section{Numerical analysis of APCW}

We adapt the technique of Ref.~\cite{Bhat2006a} suited to quantizing the electromagnetic field in the presence of dielectric media. Given a classical electric field mode $E_k(\bfr,\omega_k)$ of Maxwell's equations, the corresponding quantum electric field operator is given by $\hat{E}(\bfr)=\hat{a}_k\tilde{E}_k(\bfr,\omega_k)+\textrm{h.c}$. Here $\tilde{E}_k$ is a rescaled version of $E_k$ satisfying the energy normalization condition $\hbar\omega_k/2=\int\,d\bfr\,\epsilon_0 \epsilon(\bfr) |\tilde{E}_k(r,\omega_k)|^2$, where $\epsilon(\bfr)$ is the dimensionless electric permittivity. Interactions between an atom at $\bfr$ and the field are described by the electric dipole Hamiltonian, $H=\hat{d}\cdot\hat{E}_k(\bfr)=\hbar g_k(\bfr)\hat{a}\sigma_{eg}+\textrm{h.c.}$, where the coupling strength $\hbar g_k(\bfr)\equiv d_{eg}\cdot \tilde{E}_k(\bfr,\omega_k)$ depends on the normalized field and atomic dipole matrix element.

While the description thus far is completely general, we now specialize to the case of interactions with a single band of a 1D photonic crystal, where $k$ is now understood to be the Bloch wavevector. By Bloch's theorem, the guided mode satisfies $E_k(\bfr,\omega_k)=u_k(\bfr)e^{ikz}$, where $u_k$ is a function with periodicity given by the lattice constant $a$. Now, taking the field profile of Eq.~(3) in the main text for the localized photonic state around an atom and performing the quantization prescription, we obtain $|\bar{g}_c|=g_{\footnotesize\textrm{cell}}\sqrt{a/L}$, where $\hbar g_{\footnotesize\textrm{cell}}$ is the atom-field coupling strength of a mode at the band edge $(k=k_0)$ based upon the normalization over a single unit cell, $\hbar\omega_{k_0}/2=\int_{\footnotesize\textrm{cell}}\,d\bfr\,\epsilon_0 \epsilon(\bfr) |\tilde{E}_{k_0}(r,\omega_{k_0})|^2$. Taking the guided mode profile obtained by numerical simulation of the APCW produces a value of $g_{\footnotesize\textrm{cell}}/2\pi=12.2$~GHz.

\section{Disorder near the band edge in a photonic crystal model}

\begin{figure}
\centering
\includegraphics{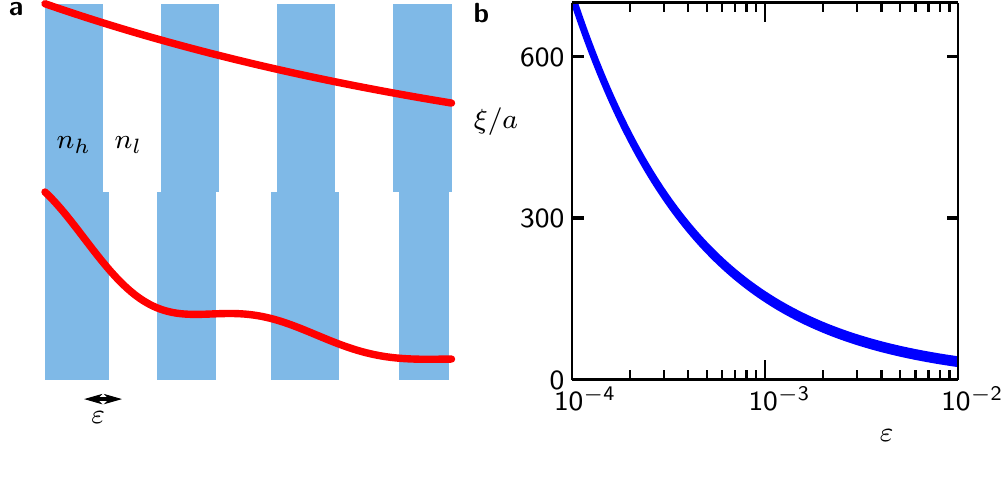}
\caption{\textbf{Effect of disorder} (a) Model of disorder in a photonic crystal consisting of alternating layers of dielectric with high $n_h$ and low $n_l$ refractive indexes. For no disorder (top half), in the band gap a field (red) decays exponentially. Introducing disorder (lower half), with standard deviation $\varepsilon$, in the phase length of each layer of the crystal may lead to faster decay of the field with a characteristic length $\xi$ and fluctuations in the intensity.
(b) Localization length $\xi$ (in units of the lattice constant $a$) as the strength of disorder $\varepsilon$ varies for $n_h/n_l=2$.}\label{fig:local}
\end{figure}

Disorder in a photonic crystal may be introduced during fabrication, either deliberately or inevitably due to the finite accuracy of the fabrication method. We model a one-dimensional photonic crystal to investigate how this disorder affects the structure of photonic modes. We take a simple model of a photonic crystal shown in Fig.~\ref{fig:local}a consisting of an alternating dielectric stack, with high refractive index $n_h$ and low refractive index $n_l$, where each layer of the stack has the same phase length $\phi_b$ at the band edge. We then introduce weak disorder by adding a random phase length shift $\varepsilon_{h(l)}\phi_b$ to each layer, which have mean $\langle\varepsilon_{h(l)}\rangle=0$, variance $\langle(\varepsilon_{h(l)})^2\rangle=\varepsilon^2$ and $\varepsilon\ll 1$.

This model of the photonic crystal, as discussed below, has the same band edge properties as the weakly disordered Kronig-Penney model, which is given by \cite{Izrailev2010a,Izrailev2012a,Derrida1984a}
\begin{equation}
\left[-\frac{\partial^2}{\partial x^2} + U\sum_n (1+\beta_n)\delta(x - a n -a\tilde{\alpha}_n)\right]\psi_q = q^2\psi_q.
\label{eq:}
\end{equation}
The Kronig-Penney model describes the propagation of the wavefunction $\psi_q$ with quasimomentum $q$ through a lattice of delta functions with spacing $a$ and strength $U$.
The disorder is introduced to the Kronig-Penney model through the random variables $\tilde{\alpha}_n$ and $\beta_n$, which adjust the position and strength of the $n$th delta function respectively, and is assumed to be weak, that is, the variables have variance $\langle \tilde{\alpha}_n^2\rangle\ll 1$ and $\langle\beta_n^2\rangle\ll 1$. This disorder in the potential leads to localization of waves in the system, which is characterized by the localization length $\xi$. In Ref.~\cite{Izrailev2010a} the stochastic dynamics of the waves in the presence of this disorder are studied by solving the associated Fokker-Planck equation, from which the localization length is found to be
\begin{equation}
\frac{\xi}{a}= \frac{2 \Gamma(1/6) }{6^{1/3}\sqrt{\pi}}\sigma^{-2/3}
\label{eq:loc_band_edge}
\end{equation}
where $\sigma$ is related to the variances of $\alpha_n=\tilde{\alpha}_{n+1}-\tilde{\alpha}_n$ and $\beta_n$ and $\Gamma$ is the Gamma function.


We now show that the model of the photonic crystal is related to the Kronig-Penney model near the band edge and as a result we can model the localization length in photonic crystals using Eq.~(\ref{eq:loc_band_edge}). While the two models appear physically different, their respective effects on wave propagation will be the same if the reflection and transmission of waves by each unit cell is the same. Indeed at the band edge,
in the absence of randomness, the models can be tuned to produce the same behavior, i.e., equal phase and magnitude of the transmission coefficient $t$, by adjusting the free parameters. As we introduce randomness we then match the models by equating the magnitude and phase of $t$ to first order in the random variables.
This gives the following relation at the band edge
\begin{align}
\beta_n &\rightarrow \frac{\phi_b (n_h - n_l) }{2\sqrt{n_h n_l}}\varepsilon_h,\label{eq:beta}\\
\alpha_n & \rightarrow \frac{\phi_b}{\phi_{KP}}\varepsilon_l  +\frac{\phi_b n_h n_l}{
 \phi_{KP}(n_h^2 - n_h n_l + n_l^2)}\varepsilon_h\label{eq:alpha}.
\end{align}

Eqs.~(\ref{eq:beta}) and (\ref{eq:alpha}) indicate that a random phase shift in the low refractive index layer of the photonic crystal is equivalent to a change in the delta function separation in the Kronig-Penney model, while a random shift in the high index layer is equivalent to changing both the delta function height and separation.
By substituting this mapping into the definition of the variance \cite{Izrailev2010a} $\sigma^2 \equiv  b^2( \phi_{KP}^2  \langle \alpha_n^2\rangle\ + \sin^2(\phi_{KP})\langle\beta_n^2\rangle)$
and utilizing $\sin^2(\phi_{KP}) =4 n_h n_l/(n_h + n_l)^2$, which holds when the band edges are matched between the two models,
we obtain
\begin{equation}
\sigma^2 =  4\phi_b^2 \left(\frac{2(r^2+1)(r-1)^2}{r(r+1)^2}
+\frac{r(r-1)^2}{(r^2-r +1)^2}
\right)\varepsilon^2
\end{equation}
where $r = n_h/n_l$ is the ratio of refractive indices. The localization length for the photonic crystal model at the band edge then follows from Eq.~(\ref{eq:loc_band_edge}).

In Fig.~\ref{fig:local}b we plot the localization length as a function of the disorder strength $\varepsilon$ for $n_h/n_l = 2$, which could be achieved with air and SiN layers for example \cite{Hung2013a}.
Here, we find localization lengths of greater than 100 unit cells, provided that the fractional standard deviation of the layer thickness is of the order of $10^{-3}$. 
For more complex geometries, simulations might be needed to quantitatively map the effects of various types of disorder onto $\sigma$, but even absent such simulations, Eq.~(\ref{eq:loc_band_edge}) serves as a useful guide as to how much disorder can be tolerated to reach a given interaction length. We note that fabrication of photonic crystals with variations on the order of $10^{-3}$ have been reported \cite{Taguchi2011a}. 


\end{document}